\documentclass[journal]{IEEEtran}

\usepackage[]{amsmath}
\usepackage{amssymb}
\usepackage{booktabs}
\usepackage{url}
\usepackage{cite}

\usepackage{tikz}
\usepackage{textcomp}
\usepackage{hyperref}

\usepackage[]{graphicx}
\graphicspath{{./figs/}}
\DeclareGraphicsExtensions{.eps,.pdf,.jpeg,.png,.jpg}

\newcommand\copyrighttext{%
  \footnotesize \textcopyright 2019 IEEE. Personal use of this material is permitted. Permission from IEEE must be obtained for all other uses, in any current or future media, including reprinting/republishing this material for advertising or promotional purposes, creating new collective works, for resale or redistribution to servers or lists, or reuse of any copyrighted component of this work in other works.}
\newcommand\copyrightnotice{%
\begin{tikzpicture}[remember picture,overlay]
\node[anchor=south,yshift=10pt] at (current page.south) {\fbox{\parbox{\dimexpr\textwidth-\fboxsep-\fboxrule\relax}{\copyrighttext}}};
\end{tikzpicture}%
}

\begin{document}

\title{Vegetation High-Impedance Faults' High-Frequency Signatures via Sparse Coding}

\author{Douglas~P.~S.~Gomes,
        Cagil~Ozansoy,~\IEEEmembership{Member,~IEEE}, and
        Anwaar~Ulhaq,~\IEEEmembership{Member,~IEEE}

\thanks{This work was supported in part by the Victorian Government Department of Economic Development, Jobs, Transport and Resources.}
\thanks{Douglas P. S. Gomes is working towards his PhD at the College of Engineering and Science, Victoria University, Melbourne, VIC 8001, Australia (email: douglas.uf@gmail.com).}
\thanks{Cagil Ozansoy is a Senior Lecturer at the College of Engineering and Science, Victoria University, PO Box 14428, Melbourne, Vic 8001, Australia (email: cagil.ozansoy@vu.edu.au).}
\thanks{Anwaar Ulhaq is Lecturer at the School of Computing and Mathematics, Charles Sturt University, NSW, Australia (email: aulhaq@csu.edu.au).}
}

\maketitle

\copyrightnotice

\begin{abstract}
The behavior of High-Impedance Faults (HIFs) in power distribution systems depends on multiple factors, making it a challenging disturbance to model. If enough data from real staged faults is provided, signal processing techniques can help reveal patterns from a specific type of fault. Such a task is implemented herein by employing the Shift-Invariant Sparse Coding (SISC) technique on a data set of staged vegetation high-impedance faults. The technique facilitates the uncoupling of shifted and convoluted patterns present in the recorded signals from fault tests. The deconvolution of these patterns was then individually studied to identify the possible repeating fault signatures. The work is primarily focused on the investigation of the under-discussed high-frequency faults signals, especially regarding voltage disturbances created by the fault currents. Therefore, the main contribution from this paper is the resulted evidence of consistent behavior from real vegetation HIFs at higher frequencies. These results can enhance phenomena awareness and support future methodologies dealing with these disturbances.
\end{abstract}


\section{Introduction}

The investigation of High-Impedance Faults (HIFs) in power distribution systems comprise an intricate field of research in the contemporary technical literature \cite{{review1},{ghaderireview}}. Despite considered as a power protection issue, these faults are low-energy events unlikely to damage power systems' equipment. Its importance is rather highlighted by the significant potential hazards to property and life. The ordinary HIF scenarios are described by an overhead power distribution conductor breaking and falling to the ground, or by the conductor's contact with nearby vegetation such as tree branches. Although important, due to the resulting hazards from having an energized conductor at ground level, the former scenario is not the focus of this paper. The investigation reported here describe findings on characteristic behaviors resulting from the contact of a power line with vegetation. The scenarios detail two categories of vegetation faults: phase to earth (conductor and earthed vegetation) and phase to phase (vegetation between conductors).

One crucial assumption made by the authors on this and past works \cite{gomes1, gomes2} is that HIFs should not be addressed as a single unified problem but rather by their distinct types. These disturbances are actually an intricate set of problems depending on many factors such as conducting surface, voltage level, network ground type, and more. The variance created by the number of variables relevant to the changes in the electrical signals makes the problem of detecting and modeling these low-energy events more challenging. In fact, works in this field can be divided into ones that propose fault detection/location schemes, and ones that intend to model the non-linear HIF characteristic impedance.

Relevant recent works \cite{{bahador}, {dos2013}, {sedighi}} on HIF models usually rely on data from real staged faults to create a variable fault impedance that mimics the effects observed in the electric signals. This, in turn, incentives methods proposing fault detection or location schemes that rely on the models to simulate fault occurrences. The proposed HIFs models \cite{dos2013 , {sedighi}} usually describe disturbances with stochastic parameters mimicking the \emph{randomness} of the faults, and other known behaviors such as build-up, intermittency, and non-symmetrical conducting cycles. 
These recent works are valuable, led to important discoveries on the generic HIF signatures, and make a real effort to separate between fault types. However, approximating these signals does not account for their whole actual signatures in the time domain. Given their sampling rate of around tens of thousand samples per second, which is on the higher end for most works, this becomes especially pertinent to the High-Frequency (HF) spectrum components. Such factor may or not be a concern for the HIFs type target in the cited papers but evidence \cite{gomes1, gomes2} shows that it has strong relevance to the faults investigated in this paper. 

The characterization of fault signatures by \textit{high-level} (hand-engineered or not) features can be useful but also ambiguous HIF descriptors. 
It may lead to a situation where detection methods can claim accurate HIF classification without clearly identifying the nature of the causal behaviours relevant in the classification process. Methods relying on high-level features in machine learning approaches are more prone to fall into this than deterministic methodologies. 
A previous work by the authors \cite{gomes1} that relied on a combination of time-frequency features and machine learning can be cited as an example. 
Despite able to accurately classify specific HIFs, the approach was unable to clearly point out the causal disturbances in recorded signal's time domain. 

In response to that, this paper investigates the use of the Shift-Invariant Sparse Coding \cite{grosse2012} technique resulted from a quest to find a powerful and effective technique to properly describe the fault patterns. This technique is able to capture patterns in the fault data, independent of its position and convolution with other signals. The resulting patterns are uncoupled and deconvoluted from each other and can have their time and frequency domain characteristics individually analysed. These capabilities are especially useful for signals of large bandwidth (hundreds of kHz) such as the ones used here. At higher frequencies, these recordings suffer from the influence of many noise sources, increasing the difficulty of associating time domain patterns with fault occurrences.  

This technique has been recently used recently in the literature. Sparse representation and neural network approach were proposed in \cite{arcsparse} to identify arc faults in distribution systems, where the learned bases resulted in a high accuracy classifier for the relevant types of arcs. In \cite{sparsedenoise}, sparse representations are used to denoise and compress power systems disturbances, that despite originated from a pre-defined set of basis, over-performs other techniques. In the field of machine fault diagnosis, sparse representations were also used to accurately classify roller bearing faults in \cite{sparsesvm} with support vector machines.

Nevertheless, this paper is novel in presenting the application of a shift-invariant version of sparse coding to learn a dictionary from a vegetation high-impedance fault data set. Such a data set differs from others in the fault detection literature especially for containing high-resolution wideband signals of these particular faults. The learned dictionary is used to create features on the signals, which are then used with a machine learning algorithm to test its relevance. The learned bases are used to further understand this fault phenomenon and its signals' features, which are validated with a linear separability calculation. Such an approach to the problem, rather than the techniques applied, is regarded as the main contribution of this paper.

The data set used in this application of sparse coding has specific characteristics that make this a particular investigation. It is made of recordings from real staged faults in a 22 kV power distribution feeder testing hundreds of vegetation samples as faults. Its signals were recorded by two channels with different band-pass characteristics: a Low-Frequency (5 Hz to 50 kHz), and an HF (10 kHz to 1 MHz). 
The currents were limited to a small energy threshold resulting in fault currents of only a few amperes. The voltages, rather than fault currents, were the domain where discriminative analysis took place. Previous works by the authors, based on this data set, proposed an HF voltage-based HIF detection method \cite{gomes1} and a comparative analysis between the information content of the LF and HF channels \cite{gomes2}. 
The fact that the latter gave evidence to the existence of relevant discriminative information to be existent only in the HF signals is the reason for focusing on this domain. 

The remaining sections of this paper present methods, patterns and consequent insights from the application of sparse coding on the data set of faults. It is hoped that it can contribute to a better understanding of the HIF phenomena and inform future detection methodologies. In short, Section 2 presents the methodology discussion providing further information on the experiments and data set characteristics. Section 3 introduces the use of the sparse coding technique for fault patterns extraction, and Section 4 presents the resulting fault signatures and inferable insights.

\section{Methodology}

\subsection{Data set characteristics}

All the findings and results presented in this paper are derived from a data set of real HIFs recordings staged on a real 22 kV distribution feeder. Given as a relevant aspect of this work, the data set of real signals were gathered in a project funded by an Australian government initiative called Powerline Bushfire Safety Program (PBSP) \cite{pbsp}. The program was created after investigations, on a series of devastating \textit{bushfires} in 2009, pointed to fault electrical assets as main causes of the most severe events \cite{vic2009}. One of the main projects, named `Vegetation Conduction Ignition Testing', was responsible for testing hundreds of vegetation samples in staged faults, creating the data set of recordings utilized in this work.

The fault tests were characterized by a few idiosyncrasies that can be summarized in the following points:
\begin{itemize}
  \item The vegetation samples were laid down between phase/earth and phase/phase conductors, creating both type of faults, with the latter representing 70\% of the distribution. 
  \item The great majority of tests had fault currents limited to 0.5 to 4 A, reflecting a particular range different from most works in the literature \cite{gomes1}. 
  \item The faults were staged in a dedicated feeder with no consumers connected to it, making most of the discriminative investigation efforts to turn from the current to the voltage signals. 
  \item The signals were sampled by two channels with different band-pass characteristics: $\thicksim$5 to 50 kHz (LF) and 10 kHz to 1 MHz (HF).   
  \item The testing feeder was part of a non-solidly grounded power distribution system. 
\end{itemize}

All these points are intricate topics that were detailed in previous works. Primarily, the authors proposed a fault detection algorithm based on the high-frequency signals in \cite{gomes1}, and then presented a comparative analysis between the fault predictor information of the LF and HF channels in \cite{gomes2}. The latter gave evidence to the existence of relevant information to discriminate between HIFs and normal states to be existent only in the HF signals. Both works focused on the voltages, rather the current signals, due to the absence of load current at the recording location and planned approach for future implementation \cite{sensor1, sensor2, gomes1}. For a more detailed analysis of the data set characteristics and sampling specifics, please refer to these works as their results were also derived from this data set.

The present analysis focuses on the staged faults' HF voltage recordings mainly due to their higher information content \cite{gomes2} when compared to LF signals. The lack of specific discussion regarding such signals in the related literature is also a relevant gap to be cited. 
One important aspect of the HF channel sampling was that, unlike the LF channel, the recordings were composed of 20-ms sweeps at every second. That is, a 20 ms sampling period followed by an interval of 980 ms with no sampling. This can be seen as a hard constraint but yet another relevant aspect of this work. As shown in \cite{gomes1}, discrimination of HIF from non-fault signals could still be successfully done with 98\% accuracy with only the available sweeps recordings.

In order to associate the identified patterns as vegetation HIF behaviors (fault signatures), separability measurements between features from the fault and non-fault signals are performed. Since these measurements are comparisons between different observations in the signal's data set, one ought to first define the meaning of \textit{observation} in this context. The recordings referred to as `fault' are 20 ms sweeps extracted immediately after the current RMS reached 0.5 A. Each fault observation is therefore extracted from a single fault test (one sweep per test) being the immediate sweep after the current reach the 0.5 threshold. The signals referred to non-fault, on the other hand, are observations extracted from recordings made in the absence of a fault in the live network feeder. These sweeps were recorded throughout the day, in the same feeder, to capture steady background signals on the days of the tests. 566 tests are used in this analysis, composing a data set of 566 fault observations which are then used in the application of the sparse coding technique. The non-fault data set also contains 566 randomly chosen sweeps from the background signals, matching the number of useful fault recordings. When separability measurements are made, the mentioned features are extracted from both fault and non-fault data sets. 

\subsection{Sparse coding}

The proposed fault pattern extraction methodology is composed of only two direct steps. The first step is responsible for extracting the fault signals' patterns with the sparse coding technique, while the second validates some of them as fault-resulting behaviors. The need for the latter is given by the fact that the network's background HF signals are not negligible at the investigated bandwidth. As a matter of fact, the recorded signals are the result of the convolution of influences from many sources: electromagnetic interference (EMI) from AM radios, non-linear loads, network transients, etc. These sources are all represented in the resulting outcomes from sparse coding indiscriminately, despite being fault-resulting behaviors or not. Hence, the task of specifically relating the found patterns to the fault occurrences needed to be addressed by a separate procedure. Such need certainly reflects some of the challenges of dealing with real sampled signals rather than synthetic data.

The methodology for sparse coding was developed as an image processing technique to characterize the primary visual cortex in mammalian receptive cells \cite{olshausen1996}. The inspiration came from the assumption that natural images have `sparse structure', i.e., they could be efficiently expressed as a small number of \textit{representations} from a \textit{larger set} of functions. The representations, in the signal processing context, are the basis functions used to describe a particular signal in a linear combination (as sinusoids in the Fourier Transform). The larger set, on the other hand, is the whole \textit{dictionary} of the possible functions used to represent the signal. The relevance of this technique to the present investigation comes mainly from its ability to learn the most efficient dictionary of bases functions to represent signals in a dataset, without any assumption about its prior distributions. These are often repeating convoluted patterns resulted from different underlying causes which can be clearly analyzed when uncoupled from one another. It is not a surprise, in this sense, that such a technique can be efficiently used to extract features in supervised learning classification tasks \cite{grosse2012}. 

The described intentions of capturing low-entropy representations can also be found in the popular dimensionality reduction method named Principal Component Analysis (PCA) \cite{WOLD198737}. In fact, both of these techniques leverage the hypothesis that the most efficient representations to describe a given data set of signals will come from the data itself. This hypothesis inspired the diversion from methods that relied on a predefined set of bases such as the Fourier or Wavelet transforms. However, differently from PCA, sparse coding does not assume that signals come from a known probability distribution \cite{olshausen1996}, making it a more adaptable but also an arduous task. Although resulting in increasing \textit{adaptability}, such generalization ability also produce a much more challenging task when learning the basis functions.
As given by (\ref{eq1}) and (\ref{eq2}), sparse coding can be described as an optimization problem over two objectives: the effectiveness of the bases at approximating the signals in a linear combination, and the sparsity of the representation. 

\begin{equation}
\begin{split}
\min_{a,s} \quad \sum_{i=1}^{m} \lVert x^{(i)} - \sum_{j=1}^{n} a^{(j)} s^{(i,j)} \rVert^{2}_{2} +  \beta \sum_{i,j} | s^{(i,j)} | \\
\end{split}
\label{eq1}
\end{equation}

\begin{equation}
s.t. \quad \lVert a^{j} \rVert^{2}_{2} \leq c, \quad 1 \leq j \leq n.
\label{eq2}
\end{equation}

The input signals $x^{(i)} \in \mathbb{R}^p, i=1,..., m$ are assumed to be a linear combinations of the dictionary of $n$ bases functions $a^{j} \in \mathbb{R}^p, j=1,..., n$ with coefficients $s^{(i,j)} \in \mathbb{R}$. $\beta$ is a positive constant that determines the trade-off between the \textit{fit} of the bases and the sparsity penalty $L_{1}$ norm. The normalization constraint in (\ref{eq2}) prevents irrelevant solutions that have too small coefficients and very large bases. This problem is not, and can't be converted to a convex problem in $s$ or $a$, meaning that it can't be directly solved in a trivial manner.

Since presented in \cite{olshausen1996}, sparse coding has proved to be a very effective representation method, gaining much attention and wide-spread use \cite{tosic2011}. However, for higher values of $p$ (dimension of the input), where patterns can appear in different parts (shifts) of a signal, this method starts to lose effectiveness. Large images where a certain object is allowed to appear at any location, or recordings where a certain speaker could start talking at any time, are examples of this. The solution proposed in \cite{grosse2012} for capturing the patterns in longer signals was to conceptualize a shift-invariant version of sparse coding. In such a version, the bases vectors can have a much smaller dimension than the input signal, which allows for the capturing of smaller shifted repeating patterns. The product of matrices $a^{(j)} s^{(i,j)}$ becomes a convolution, $a^{(j)}$ assumes a dimension smaller than $p$, let it be $q$, and the coefficients becomes vectors $s^{(i,j)} \in \mathbb{R}^{p-q+1}$, as showed in (\ref{eq3}) and (\ref{eq4}).

\begin{equation}
\begin{split}
\min_{a,s} \quad \sum_{i=1}^{m} & \lVert x^{(i)} -  \sum_{j=1}^{n} a^{(j)} \ast s^{(i,j)} \rVert^{2}_{2} \\ +   & \beta \sum_{i,j}  \lVert s^{(i,j)} \rVert_{1}
\end{split}
\label{eq3}
\end{equation}

\begin{equation}
s.t. \quad \lVert a^{j} \rVert^{2}_{2} \leq c, \quad 1 \leq j \leq n.
\label{eq4}
\end{equation}

An efficient solution for the problem in (\ref{eq1}) and (\ref{eq2}) was proposed in \cite{lee2007}. The authors made use of the fact that although not convex in $s$ and $a$ simultaneously, the optimization problem becomes convex if any of these parameters are considered individually. The two-step proposed algorithm first assumes the bases to be constant vectors while it optimizes over the coefficients. Then, it does the same with coefficients, freezing their values as it optimizes the bases. 
Convergence is reached by the iterative optimization of these two consecutive steps until the objective function stops decreasing. The solution for the shift-invariant problem was proposed by \cite{grosse2012} where the challenging problem of optimizing the convolved basis was solved by extending the algorithms proposed by \cite{lee2007}. The extension included mathematical manipulations, like the translation of the variables to the frequency domain to solve a Lagrange Dual problem (convolution as multiplication), and further efficient ways to solve for a large number of coefficients resulting from the new problem framing.

In the application of this technique, there are some relevant hyperparameters to consider. The most relevant one is the number of bases in the dictionary, the length of the basis functions, and the beta regularization value. The first two are explored in more detail in the results section. The regularization value, however, was set empirically to a value that allowed the dictionary to be learned in a period of 12 to 48 hours. As it represents the trade-off between fitness and sparsity, a high beta value would lead to more iterations, unnecessarily expanding the computational time without much relevancy to final results. Please refer to the works of Grosse et al. \cite{grosse2012} and Lee et al. \cite{lee2007} for more detailed information on the application and hyperparameters.

\subsection{Basis validation}

The application of sparse coding will capture all patterns (fault transients, EMI, and noises) indiscriminately. This means that a procedure to measure the relation of the resulting basis or dictionary with fault occurrences needs to be introduced. The proposed method used herein relies on the existence of labelled data from fault and non-fault signals to calculate a measurement of variance between the dictionary, or basis tested, with the two possible labels. 

When testing dictionaries, the measurement is given by the resulted accuracy from a machine learning classification approach, with features extracted from the signals in the data set. Inspired by the theory of Convolution Neural Networks (CNNs), the extracted features are the result of using the patterns as filters in a cross-correlation operation, measuring its similarity to the input signal at all possible shifts. The convolved signal is then non-linearly summed, as passed through a rectified linear unit (ReLU) activation layer, and fed to the classifier as a similarity feature. Each base function is responsible for creating one feature on all signal,s i.e., each signal will have the same number of features as the number of bases in the tested dictionary. The machine learning technique is an off-the-shelf ensemble of decision trees \cite{freund1996} which is validated by 10-fold cross-validation. Methods such as the random forest or boosted decision trees are powerful discriminators, have a relatively simple implementation, and can be easily adjusted to avoid overfitting with hyperparameters like the maximum number of splits or maximum leaf size.

To label an identified pattern as a fault signature, however, a relation score for each basis on a dictionary regarding fault occurrences must be calculated. The proposed method here, in a similar approach to the dictionary evaluation, use the discriminative power from the individual features as a measure of the variance between classes. At this step, instead of learning a classifier with many features, each basis is scored by the resulted separability from a single linear split separator on that related feature. This separability index is based on the Gini Impurity (GI), a measure used in the CART (Classification and Regression Tree) algorithm \cite{breiman1984} to evaluate a binary decision boundary regarding its information gain. In CART, the GI is used to decide where to split the nodes of trees by the information gain given by each potential split. It often considers the different ways of creating the decision boundary in each feature and selects the one with the smallest GI (highest information gain). The way it was adopted in this investigation was to consider every data point in the tested feature as a potential split. The basic calculation is given in (\ref{gieq}), where the result is the summation the probability $p_{i}$ of an item with label $i$ from $J$ number of labels being chosen, times the probability of mistakenly categorizing that item. When all the observations have their GI calculated, the algorithm selects the smallest one to be the separability index for that feature. If the feature has high invariance (high separability index), one can conclude that it is a good high-level representation of the consequences of a fault occurrence. As each feature is associated with a single basis function, the ones that presented this higher separability potential are then be studied as a recurring pattern in fault event (fault signature) where many valuable conclusions can be derived.

\begin{equation}
GI=\sum_{i=1}^J  p_i \sum_{k\neq i} p_k=1-\sum_{i=1}^J p_i^2
\label{gieq}
\end{equation}

The methodology algorithm can be summarized:
\begin{itemize}
\item First part
    \begin{enumerate}
    \item Get the database of signals.
    \item Select relevant fault observations (sweeps).
    \item Use fault observations as input to the shift-invariant sparse coding algorithm setting hyperparameters:
    \begin{itemize}
		\item Number of basis    	
    	\item Length of basis
    	\item Iterations
    	\item Batch size
    	\item Regularization parameter
    \end{itemize}
    \item Get the returned dictionary and use the bases to create fault and non-fault observations features.
    \item Use the features to learn a Random Forest classifier.
    \item Validate the classifier with 10-fold cross-validation
    \item Repeat from step (3) with different relevant hyperparameters
    \end{enumerate}

\item Second part
    \begin{enumerate}
    \item Get the dictionary with highest accuracy.
    \item Create fault and non-fault observations features.
    \item Calculate the separability via G.I. for each feature/basis to attest for invariance between classes 
    \end{enumerate}
\end{itemize}

\section{Results}

\subsection{Captured patterns}

\begin{figure}[t]
\centering
\includegraphics[scale=0.5]{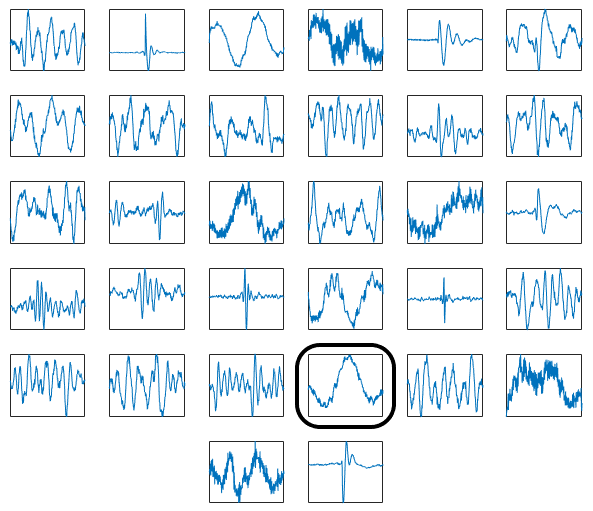}
\caption{Example of a learned 32-basis dictionary.}
\label{32_basis}
\end{figure}

To exemplify the outcome of the sparse coding algorithm, Fig. \ref{32_basis} presents the returned learned dictionary when 32 is set as the number of bases, performed over 100 iterations, spanning the whole fault signals dataset. 
This length represents a duration of 125 $\mu$s, which is a bit more than the period of the lowest considered frequency (10 kHz), limited by the sampling channel high-pass filter. As a matter of fact, the narrow frequency band close to 10 kHz was intensely present in the HF signals. Bases such as the one shown in the fourth column, fifth row, of Fig. \ref{32_basis} are an example of this.

\begin{figure}[t]
\centering
\includegraphics[scale=0.7]{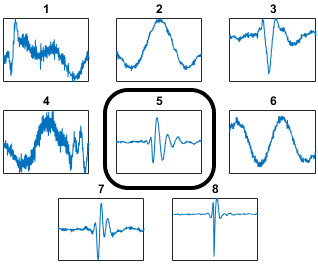}
\caption{Learned 8-basis dictionary.}
\label{8_basis}
\end{figure}

As seen in Fig. \ref{32_basis}, when the number of 32 functions is chosen as a hyperparameter, there are a considerable number of redundant bases trying to describe a similar pattern. This redundancy suggests that the number of possible underlying sources creating these patterns may not be so numerous. This is somewhat expected from a 50 Hz power distribution system at higher frequencies. Nevertheless, having this redundancy in the signal representation is possibly not advantageous when trying to narrow down the fault signature representation with the present validation technique. The underlying hypothesis is that the information to discriminate between fault and non-fault signals will be ``diluted'' in the redundant high-level representations (features), unnecessarily increasing the complexity of the statistical model. If such a hypothesis is right, the number of bases functions becomes an important hyperparameter to consider. This was tested by creating dictionaries with a different number of bases and testing their resulting features' discriminative potential. The results presented by cross-validation in Table \ref{nbasistab} suggest this hypothesis to be right. In fact, it also suggests that, in terms of the relation between separability and number of bases functions, less is better. The returned 8-basis dictionary, which resulted in higher separability and now used for the remaining discussions in this paper, is illustrated in Fig. \ref{8_basis}.

\setlength{\heavyrulewidth}{1.5pt}
\setlength{\abovetopsep}{4pt}
\begin{table}[t]
\caption{Discriminative potential vs. number of basis functions}
\label{nbasistab}
\centering
\resizebox{\columnwidth}{!} {
\begin{tabular}{*4c}
\toprule 
No. of functions & \bfseries Accuracy (\%) &  True positives (\%) & True negatives (\%)\\
\midrule
 8 & 94.52 & 92.4 & 96.64\\
 16 & 94.08 & 91.52 & 96.64\\
 32 & 93.37 & 89.93 & 96.82\\
 64 & 90.11 & 85.51 & 94.70\\
128 & 88.69 & 82.69 & 94.70\\
\bottomrule
\end{tabular}
}
\end{table}

As shown in Table \ref{nbasistab}, the 8-basis dictionary can correctly discriminate between fault and non-fault classes with more than 94\% accuracy. This is especially relevant since the features used are given by simple cross-correlation calculations between the basis and signals. 

Associating a particular basis function as a fault signature, however, requires a more detailed description of the \textit{effectiveness} of each pattern. This was done by creating a simple linear decision boundary in the feature resulted from each singular basis. Instead of using all the basis-resulting features to learn an ensemble of decision trees, this approach used only the one-dimensional data related to each feature to create a one-split linear separator. The result, listed in descending order from the most to less effective basis, is shown in Table \ref{indbasistab}. 

\setlength{\heavyrulewidth}{1.5pt}
\setlength{\abovetopsep}{4pt}
\begin{table}[t]
\caption{Individual discriminative potential of each basis listed in descending order.}
\label{indbasistab}
\centering
\begin{tabular}{*2c}
\toprule 
Functions number & Separability (\%) \\
\midrule
 5 & 90.46 \\
 8 & 88.78 \\
 7 & 86.48 \\
 3 & 77.74 \\
 2 & 63.52 \\
 6 & 63.16 \\
 4 & 62.90 \\
 1 & 59.72 \\
\bottomrule
\end{tabular}
\end{table}

The results presented in Table \ref{indbasistab} not only show that function no. 5 does most of the work when distinguishing between classes (reasons further discussed) but also points to an important aspect of fault signals. Note that the patterns shown in Fig. \ref{8_basis} can actually be separated in functions that tried to fit the stationary sinusoidal components such as no. 2, 4, and 6, and others resulted from fitting transients with finite, short existence. This, together with the results shown by Table \ref{indbasistab}, suggests that the effect of a vegetation HIF in the HF signals is mostly given by an added transient component. 

\subsection{Pattern analysis}

\begin{figure}[t]
\centering
\includegraphics[scale=0.8]{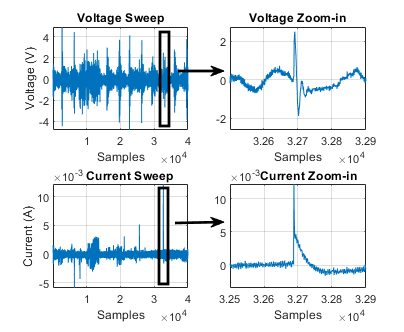}
\caption{Example of a in-fault first Voltage and Current sweeps zoomed in at strong HF current transient.}
\label{trans_ex}
\end{figure}

\begin{figure}[t]
\centering
\includegraphics[scale=0.65]{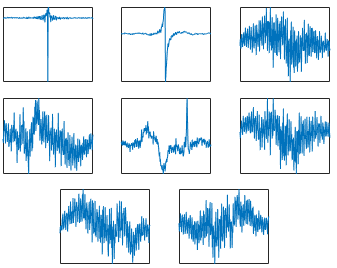}
\caption{8-basis learned dictionary from the current signals.}
\label{8basis_current}
\end{figure}

The simplest explanation for the transient events is related to the second-order circuit response to impulses or steps in the HF current signals. Examples of this can actually be found when peaks of the cross-correlation are followed, as illustrated by Fig. \ref{trans_ex}. Nonetheless, although found in the signals, such examples are not common or easily located. There are a few reasons for this: the current transients are rarely that strong and isolated, the peaks created in the voltage HF signals are usually smaller than background noise (as in the image), and their appearance in time are stochastic (shifted and non-deterministic). HF current bursts are often given by isolated or shifted convoluted discontinuities buried in noise, such as the ones shown in Fig. \ref{trans_ex} around the 12k\textsuperscript{th} sample on the bottom left plot. Fig. \ref{8basis_current} illustrates an 8-basis dictionary learned from the HF current signals with the same hyperparameters used to create previous dictionaries.

Another mentioned hyperparameter is the length of the basis function in the dictionary or, more simply, their number of samples. All the results showed here consider 250 samples, or 125 us in duration, as the bases' length. The reason for doing so was that no relevant differences in results were found when varying the length anywhere from 25 to 500 samples. The illustrated patterns, in this manner, had the length with more convenient visualization aspects.

Despite the mentioned challenges, tools such as sparse coding can reveal these hidden patterns, which can aid in understanding the vegetation HIF phenomena. In fact, this methodology was first adopted as an attempt to pinpoint the reasons why the fault detection method, previously proposed by the authors \cite{gomes1}, obtained 98\% accuracy at such a task. The former methodology also used simple features with an ensemble of decisions trees to create a signal classifier. However, the previous features were given by the coefficients of a wavelet transformation using the fourth order Symlet as the mother wavelet. Although relevant dependability and security were presented, the work could not narrow down the reason why the Symlet wavelet was so effective. This is a problem especially relevant for methodologies using machine learning techniques as classifiers. 

\subsection{Wavelet similarities}

To better understand the past methodology's effectiveness, two important points are worth remembering: (1) the coefficients of the discrete wavelet transform is basically the result of the iterative convolution of the scaling and mother wavelet filters with the downsampled signals; and (2) the convolution of two signals is equal to the element-wise product of these signals in the frequency domain (Parseval's Theorem). This means that the wavelet coefficients, which reflect the signal's energy at different bandwidths, 
can also be given by the element-wise product of the signal and related filter in the frequency domain. That is, the wavelet coefficients features are the result of the iterative application of filter banks with different band-pass characteristics. The features are then extracted by a sum operator (linear or not) on these coefficients, in the same manner as the features explored here by cross-correlation. It may be worth remembering that the convolution and cross-correlation only differ in direction of the product-sum operations, which disappear when they are all summed up in high-level features. 

\begin{figure}[t]
\centering
\includegraphics[scale=0.75]{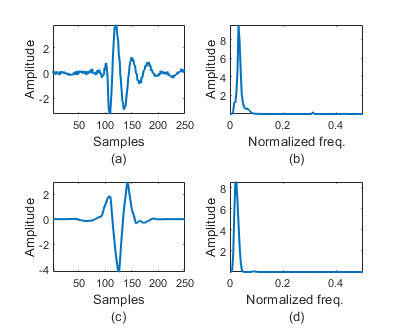}
\caption{Comparison between basis function and wavelet. Basis no. 5 in a) time domain and b) frequency domain. Symlet level 3 in c) time domain and d) frequency domain.}
\label{basis_vs_wavelet}
\end{figure}

The reasons for the effectiveness of the wavelet transform, therefore, was most evident with a comparison between the most efficient basis (no. 5) and the third level Symlet (`sym4') in the frequency and time domain, illustrated by Fig. \ref{basis_vs_wavelet}. Moreover, when a simple 4-level Symlet wavelet decomposition of order 4 is used in conjunction with the features presented here the learn a classifier, accuracies result are often higher than 98\%.

\subsection{Applicability and Significance}

Despite explaining the most variance between fault and non-fault signals, basis no. 5 with 90.46\% and level 3 Symlet with 94.7\% separability, there are still important considerations regarding the fault signatures.  If these transients are indeed second-order responses from current transients, they would considerably change given different RLC parameters. This may be taken as a strong utility constraint of the herein presented findings. However, as seen in the 32-basis dictionary in Fig. \ref{32_basis}, even when the same test rig, network, and fault distance are considered, the response can still drastically vary due to the different resistance and inductance of the contact surface (vegetation sample). Despite such variance, nevertheless, the results presented by the features created from sparse coding filters and the wavelet transform are still efficient at discriminating fault from non-fault signals. The wavelet transform, in specific, is a powerful tool for the task of fault detection since the filters' frequency response is smooth and can capture a well-defined tuned bandwidth of resonant transients created in the HF signals. That is, even if the RLC parameters change considerably, the coefficients would still be relevantly activated and useful for detection. This result may be carried over through the field in an electric grid contemplating distributed measurements where the line parameters are bounded or detailed understood. Additionally, this can serve as a counter-argument which other works will have to address when trying to dismiss the use of the wavelet transform, given that it is highly effective and not extremely demanding in computational complexity, when proposing a novel and more complex technique. 

Finally, the significance of such findings and possible implementation issues must be mentioned. For example, although the application of Sparse Coding to find recurring patterns was successfully followed, even the most effective basis has a relevant presence in both non-fault and fault signals. This is the reason why the separability potential of none of the features reached 100\%. What is significant, nevertheless, is that the value of this feature is often tens of times higher in fault signals, meaning that they are much more recurrent in fault occurrences. Readers familiar with the ``Partial Discharge" literature, which often quantify such responses with reflectometry, would not be surprised with the results presented herein. Power distribution system HIFs are often discussed as random events with characteristics hard to quantify such as build up, non-linearity, and high intermittency \cite{review1, ghaderireview}. The specificity and quantitative evidence presented here, therefore, can add to the phenomena understanding, especially regarding the high-frequency domain components. Differently from originally thought, the appearances of these high-frequency patterns are not totally random. A great part of them appears in zero crossings in the voltage power frequency cycle, as results from the non-linearity of conduction in the vegetation surface. The sudden conduction which starts only after the voltage reaches a certain threshold, like in an anti-parallel diode branch, is the main cause of this non-linearity. 

The potential application of the insights given by the sparse coding in the field of fault detection is also relevant. When comparing to the methodology proposed in \cite{gomes1}, such patterns could be added as additional features increasing the method's dependability and security. Regarding the constraints, the primary challenges of such an application would definitely be the sampling rate necessary to capture the high-frequency signals. Although a serious obstacle, some evidence points it to be a manageable constraint not to be fastly disregarded. Novel sampling technologies \cite{sensor1,sensor2} and increasing accessibility of computational power \cite{rasp} will make the application of such methods more plausible in a modern smart grid scenario, with probably more distributed and detailed measurements. As a matter of fact, future works by the authors will contemplate a low-cost prototype which is now moving out of the proof-of-concept phase. 

Despite the challenging task, the authors argue for the need for enhanced methods if vegetation HIFs of low fault currents (from 0.5 A) are to be detected (especially in non-solidly grounded systems). The tests responsible for the utilized data set, for example, had a commercial relay with embedded HIF detection function in the testing feeder that did not detect any of the staged faults \cite{marxsen}. The discriminative factor, moreover, was only to be found when investigating the high-frequency signals \cite{gomes2}, appearing to play a critical role despite the utilized detection methodology.

\section{Conclusions}
The experiments and results presented herein attest for the powerful sparse coding capabilities at capturing shifted patterns in signals' time domain. When applied to the fault signals data set, the redundancy presented at dictionaries of different sizes also revealed that the number of sources creating these patterns may not be so numerous. Fortunately, through the adopted validation method, some of the captured patterns were highly correlated to fault occurrences. These were then labelled and analysed as \emph{fault signatures}. 

The patterns in the voltage signals point them to be a response to the isolated or convoluted, high-frequency discontinuities in the fault current. Moreover, a comparison between the most useful patterns and the Symlet filter revealed the reasons why the wavelets were so effective at discriminating fault occurrences in previous works. The results by cross-validation also help to attest the robustness of similar techniques when used to discriminate non-fault and fault signals. They showed relevant predictor potential even when considerable changes in the parameters responsible for the transients' behaviour were present throughout the tests. 

The authors believe that such insights can serve as quantitative evidence of the presence of vegetation HIFs fault patterns, which can enhance phenomena understanding, and possibly aid future detection/classification events methodologies.

\section*{Acknowledgement}
The authors would like to thank the Powerline Bushfire Safety Program and the Department of Economic Development, Jobs, Transport and Resources from Victoria – Australia for funding such massive program and making data from the tests available. 

\bibliographystyle{IEEEtran}

\bibliography{IEEEabrv,references}

\vspace{-15 mm}
\begin{IEEEbiography}[{\includegraphics[width=1in,height=1.25in,clip]{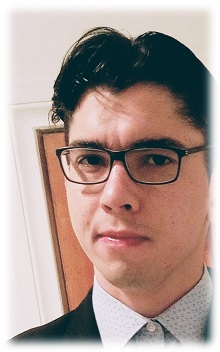}}]{Douglas P. S. Gomes}
received the B.E. degree in electrical engineering from Federal University of Mato Grosso, Brazil, in 2013 and the M.Sc. degree in Power Systems from University of Sao Paulo, Brazil, in 2016. He is currently working towards his Ph.D. degree at Victoria University, Melbourne, Australia. His research interests are power systems, protection, power quality, and artificial intelligent systems.
\end{IEEEbiography}

\vspace{-15 mm}
\begin{IEEEbiography}[{\includegraphics[width=1in,height=1.25in,clip]{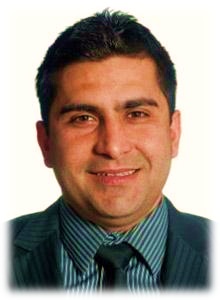}}]{Cagil Ozansoy}
received his B.Eng. degree in electrical and electronic engineering (Hons.) and the Ph.D. research degree in power system communications from Victoria University, Melbourne, Australia, in 2002 and 2006, respectively. He is now working as a Senior Lecturer and Researcher in the College of Engineering and Science, Victoria University. His major teaching and research focus is in electrical engineering, renewable energy technologies, energy storage, and distributed generation. He has successfully carried out and supervised many sustainability-related studies in collaboration with local governments in the past. He has over 25 publications detailing his work and contributions to knowledge.
\end{IEEEbiography}

\vspace{-15 mm}
\begin{IEEEbiography}[{\includegraphics[width=1in,height=1.25in,clip]{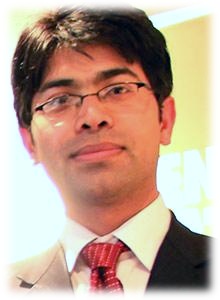}}]{Anwaar Ulhaq}
received the B.S.(Hons) in computer software Engineering from University of Engineering and Technology, Lahore, Pakistan. He received M.S. System Engineering degree from G.I.K institute, Topi, Pakistan and Ph.D. (Intelligent Systems) from Monash University, Australia. He is working as a lecturer, information technology in College of Engineering and Science, Victoria University. His research interest includes pattern recognition, signal and image processing and human action recognition.
\end{IEEEbiography}

\end{document}